\documentclass[prd,preprintnumbers,showkeys,nofootinbib]{article}
\usepackage{bm, graphicx,amsmath,amssymb,epsf}

\usepackage{graphicx}

\begin{document}

\title{\Large \bf  Duality symmetry of  BFKL equation: reggeized gluons vs color dipoles. }
\author{\large Alex~Prygarin
\bigskip \\
{\it  II. Institute for Theoretical Physics, Hamburg University, Germany} \\
 }
\date{}
\maketitle

\begin{abstract}

  We show that the duality symmetry of the BFKL equation can be interpreted as a symmetry under rotation of the BFKL Kernel in the transverse space from $s$-channel~(color dipole model) to $t$-channel~(reggeized gluon formulation).  We argue that the duality symmetry holds also in the non-forward case due to a very special structure of the non-forward BFKL Kernel, which can be written as a sum of three forward BFKL Kernels. 
The duality symmetry is established by identifying the dual coordinates with the transverse coordinates of a non-diagonal dipole scattered off the target.

\end{abstract}

\section{Introduction}
\label{sec:intro}

\hspace{0.4cm} The Balistky-Fadin-Kuraev-Lipatov~(BFKL) equation \cite{BFKL} describes the amplitude of scattering at very high center-of-mass energy $\sqrt{s}$ with $|t/s|\ll 1 $, where $t$ is the square of the transferred momentum. The leading order BFKL is obtained by summing terms $(\alpha_s \log s)^n$, where each power of the coupling constant $\alpha_s$ is accompanied by the corresponding power of the logarithm of energy. This kinematic regime  is called multiregge kinematics. In the multiregge kinematics, the transverse degrees of freedom fully decouple from the longitudinal ones. This allows to formulate the BFKL equation as evolution in complex time~(rapidity) with the integral Kernel operating in the transverse space.  The BFKL equation was originally formulated using the fact that $t$-channel gluons reggeize and the production vertices of the $s$-channel gluons factorize in the Regge kinematics.
In this picture the BFKL equation describes a compound state of two reggeized gluons. 
An alternative derivation of the BFKL evolution was proprosed by Mueller \cite{MUCD} using $s$-channel unitarity for evolution of colorless dipoles in the limit of the large number of colors. 
The BFKL equation was solved \cite{BFKLsol} using the conformal invariance of the BFKL Kernel. It was also noticed that the BFKL Kernel has another interesting property called  the \emph{duality} symmetry found by Lipatov~\cite{Lipatov:1998as}. This symmetry  means that the form of the BFKL equation does not change if  the gluon momentum $k$ is replaced by its conjugate coordinate. It was shown that this symmetry can explain the integrability of the BFKL equation. However, it was also suggested  that the duality symmetry should hold only for the case of zero momentum transfer  for a system of two reggeized gluons.

The objective of the present study is to show that the duality symmetry of the BFKL equation holds also in the non-forward case, though not in an explicit way. We continue the analysis started by the author \cite{Prygarin:2009tn} and establish the duality symmetry as a symmetry between reggeized gluon formulation and the dipole picture of the BFKL evolution. In particular,  we show that the evolution equation for a dipole with different sizes to the left and to the right of the unitarity cut can be written in the form of the BFKL equation in the dual coordinates. The dual momenta coordinates and the conjugate coordinates are not \emph{a priori} related objects, the fact that can identify them is to be understood as a sign for the duality symmetry. However, there seem to be no obvious choice of the Fourier transform (at least of a single variable) that can take one picture to another. This is the reason why we prefer to call this symmetry - the \emph{hidden} duality symmetry.  The hidden duality symmetry can also be interpreted as a symmetry under rotation of the BFKL Kernel in the transverse space from $t$-channel~(reggeized gluons) to $s$-channel~(color dipoles) and back.

\section{Duality symmetry of BFKL equation}
\label{sec:dualityBFKL}
\hspace{0.4cm} In this section we explain briefly how the duality symmetry appears in the leading order BFKL equation and show why it is related to the integrability.  

The duality symmetry of the system of interacting reggeons in the limit of a large number of colors was first formulated by Lipatov~\cite{Lipatov:1998as}. In the following we briefly outline the major relevant points of this study. 

 We start with a general description of the BFKL approach and present its formulation in  terms of the holomorphic Hamiltonian in the Sch\"odiner like equation.

\begin{figure}[h]
\begin{center}
\includegraphics[width=1.1in]{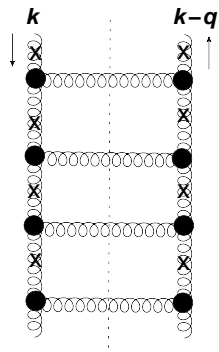}  
\end{center}
\caption{The BFKL evolution equation describes high energy scattering as a compound state of two $t$-channel reggeized gluons, with real $s$-channel gluon emissions crossing the dashed line of the unitarity cut. The effective real production vertices are denoted by the dark blobs and the fact that $t$-channel gluons are reggeized is reflected by crosses.  } \label{fig:bfkl}
\end{figure}

The BFKL equation describes the behavior of the scattering amplitude in the limit of the center-of-mass energy $\sqrt{s}$ being much larger than the typical transferred momentum $|t/s|\ll 1$~(the Regge kinematics). The leading order BFKL evolution equation is obtained by summing the powers of the parameter $\alpha_s \log s$, where each power of the strong coupling constant $\alpha_s$ is accompanied by the corresponding power of the logarithm of energy. In this picture the BFKL Pomeron appears as a compound state of two reggeized gluons of transverse momenta $\vec{k}$ and $\vec{k}-\vec{q}$ as illustrated in Fig.~\ref{fig:bfkl}. The color singlet BFKL in the limit of large number of color $N_c$ reads

\begin{eqnarray}\label{bfkl}
\left(\frac{\partial }{\partial y}-\epsilon(-\vec{k}^2)-\epsilon(-(\vec{k}-\vec{q})^2)\right)\mathcal{F}(\vec{k},\vec{k}-\vec{q})=
\frac{\alpha_s N_c}{2\pi^2} \int d^2 \vec{\chi} \frac{K(\vec{k},\vec{\chi})}{\vec{\chi}^2 (\vec{\chi}-\vec{q})^2}\mathcal{F}(\vec{\chi},\vec{\chi}-\vec{q})
\end{eqnarray}
where the gluon reggeization enters the equation through the Regge gluon trajectory
\begin{eqnarray}\label{regge}
\epsilon(-\vec{k}^2)=\frac{\alpha_s N_c}{4\pi^2} \int d^2 \vec{\chi} \frac{-\vec{k}^2}{\vec{\chi}^2(\vec{\chi}-\vec{k})^2}
\end{eqnarray}

The real emission part of the Kernel is given by 
\begin{eqnarray}\label{kernel}
K(\vec{k},\vec{\chi})=\vec{q}^2-\frac{\vec{k}^2(\vec{\chi}-\vec{q})^2}{(\vec{\chi}-\vec{k})^2}-\frac{\vec{\chi}^2(\vec{k}-\vec{q})^2}{(\vec{\chi}-\vec{k})^2}
\end{eqnarray}

In the leading order BFKL the transverse momenta components decouple from the longitudinal ones~(rapidity).  Due to this factorization the BFKL Pomeron can be written as a state in the two dimensional transverse space that evolves with rapidity which plays a role of an imaginary time. This fact makes it possible to formulate the color singlet BFKL dynamics in the form of the Sch\"odinger equation for the wave function $f_{m,\tilde{m}}(\vec{\rho}_1,\vec{\rho}_2,...,\vec{\rho}_n;\vec{\rho}_0)$ for a system of $n$-reggeized gluons \cite{Kwiecinski:1980wb,Bartels:1980pe,Lipatov:1990zb}, the BFKL equation is obtained for $n=2$.
The vectors $\vec{\rho}_k$ are two dimensional coordinates of the reggeized gluons, and $m$ and $\tilde{m}$ are  the conformal weights 
\begin{eqnarray}
 m=\frac{1}{2}+i\nu+\frac{n}{2} ,\;\;\; \tilde{m}=\frac{1}{2}+i\nu-\frac{n}{2}
\end{eqnarray}
 which are expressed in terms of the anomalous dimension $\gamma=1+2i\nu$ and the integer conformal spin $n$. The anomalous dimension and the conformal spin in this context were introduced when solving the BFKL equation in the complex coordinates 
\begin{eqnarray}
 \rho_k=x_k+iy_k, \;\;\; \rho^*_k=x_k-iy_k 
\end{eqnarray}
using the conformal properties of the BFKL Kernel. 

The BFKL wave function $f_{m,\tilde{m}}$ satisfies the Sch\"odinger equation
\begin{eqnarray}
 E_{m,\tilde{m}}f_{m,\tilde{m}}=H f_{m,\tilde{m}}
\end{eqnarray}
with the energy $E_{m,\tilde{m}}$ being proportional to the position of the singularity in the complex angular momentum $j$ plane. In the multicolor limit the Hamiltonian possesses a property of holomorphic separability 
\begin{eqnarray}
 H=\frac{1}{2}\left(h+h^*\right)
\end{eqnarray}
where the holomorphic and the anti-holomorphic Hamiltonians 
\begin{eqnarray}\label{hamsmall}
 h=\sum_{k=1}^n h_{k,k+1}, \;\;\; h^*=\sum_{k=1}^n h^*_{k,k+1}
\end{eqnarray}

are expressed through the BFKL operator \cite{Lipatov:1993qn}
\begin{eqnarray}\label{hholom}
 h_{k,k+1}=\log(p_k)+\log(p_{k+1})+\frac{1}{p_k}\log(\rho_{k+1})p_k+\frac{1}{p_{k+1}}\log(\rho_{k+1})p_{k+1}+2\gamma
\end{eqnarray}
In Eq.~(\ref{hholom}) one defines $\rho_{k,k+1}=\rho_k-\rho_{k+1}$, $p_k=i\partial/(\partial \rho_k)$, $p^*_k=i\partial/(\partial \rho^*_k)$ and $\gamma=-\psi(1)$~(the Euler constant). The holomorphic separability of the Hamiltonian means the holomorphic factorization  of the wave function 
\begin{eqnarray}\label{confbfkl}
f_{m,\tilde{m}}(\vec{\rho}_1,\vec{\rho}_2,...,\vec{\rho}_n;\vec{\rho}_0)=\sum_{r,l} c_{r,l}f^r_{m}(\rho_1,\rho_2,...,\rho_n;\rho_0)
f^l_{\tilde{m}}(\rho^*_1,\rho^*_2,...,\rho^*_n;\rho^*_0)
\end{eqnarray}
 and the Sch\"odinger equations in the holomorphic and the anti-holomorphic spaces 
\begin{eqnarray}
 \epsilon_m f_m=h f_m, \;\;\; \epsilon_{\tilde{m}}f_{\tilde{m}}=h^* f_{\tilde{m}}, \;\;\;E_{m,\tilde{m}}=\epsilon_m+\epsilon_{\tilde{m}}
\end{eqnarray}
The degenerate solutions are accounted for by the coefficients $c_{r,l}$ in Eq.~(\ref{confbfkl}), which are fixed by the singlevaluedness condition for the wave function in the two dimensional space.

It is interesting to note that the BFKL way function can be normalized in two different ways
\begin{eqnarray}
 \parallel f\parallel^2_1=\int \prod_{r=1}^n d^2 \rho_r \left| \prod^n_{r=1} \rho^{-1}_{r,r+1} f\right|^2,
\;\;\;
 \parallel f\parallel^2_2=\int \prod_{r=1}^n d^2 \rho_r \left| \prod^n_{r=1} p_{r} f\right|^2
\end{eqnarray}
This is in an agreement with the hermicity properties of the Hamiltonian, since the transposed Hamiltonian $h^t$ can be obtained by two different similarity transformations   \cite{Lipatov:1993yb}
\begin{eqnarray}\label{norm}
 h^t=\prod_{r=1}^n p_r h \prod_{r=1}^n p_r^{-1}=\prod_{r=1}^n \rho_{r,r+1}^{-1} h \prod_{r=1}^n \rho_{r,r+1}
\end{eqnarray}

The BFKL Hamiltonian is invariant under cyclic permutations corresponding to the Bose symmetry of the reggeon wave function $i \to i+1 \;(i=1,2...,n)$ in 
multicolor limit. It was noticed by Lipatov~\cite{Lipatov:1998as} that the Hamiltonian is also invariant under canonical transformation 
\begin{eqnarray}\label{change}
 \rho_{k-1,k} \to p_{k} \to \rho_{k,k+1}
\end{eqnarray}
accompanied by the change of the operator ordering. This property becomes obvious if we rewrite the Hamiltonian Eq.~(\ref{hamsmall}) in the form of 
\begin{eqnarray}
 h=h_p+h_{\rho}
\end{eqnarray}
with
\begin{eqnarray}\label{hp}
h_p=\sum_{k=1}^n \left( \log(p_k) +\frac{1}{2} \sum_{\lambda=\pm1}  \rho_{k,k+\lambda} \log(p_k) \rho^{-1}_{k,k+\lambda}+\gamma \right)
\end{eqnarray}

and 
\begin{eqnarray}\label{hrho}
h_\rho=\sum_{k=1}^n \left( \log(\rho_{k,k+1}) +\frac{1}{2} \sum_{\lambda=\pm1}  p^{-1}_{k+(1+\lambda)/2} \log(\rho_{k,k+1}) p_{k+(1+\lambda)/2}+\gamma \right)
\end{eqnarray}

The invariance of the BFKL Hamiltonian under the change of the variables Eq.~(\ref{change}) together with  the change of the operator ordering was called the \emph{duality} symmetry. 
The duality symmetry   implies that the BFKL Hamiltonian commutes $[h,A]=0$ with the differential operator
\begin{eqnarray}
 A=\rho_{12}\rho_{23}...\rho_{n1}p_1p_2...p_n.
\end{eqnarray}
 or, more generally, there is a family of mutually commuting integrals of motion \cite{Lipatov:1993yb}
\begin{eqnarray}
 [q_r,q_s]=0, \;\;\; [q_r,h]=0
\end{eqnarray}
and they are given by 
\begin{eqnarray}
 q_r=\sum_{i_1<i_2<...<i_r} \rho_{i_1,i_2}\rho_{i_2,i_3}...\rho_{i_r,i_1}p_{i_1}p_{i_2}... p_{i_r}
\end{eqnarray}
The operators $q_r$  build a complete set of the invariants of the transformation. Therefore the Hamiltonian $h$ is their function
\begin{eqnarray}
 h=h(q_2,q_3,...,q_n)
\end{eqnarray}
and a common eigenfunction of $q_r$ is simultaneously a solution to the Sch\"odinger equation. This fact explains why the duality symmetry is related to the integrability of a system of Reggeons in the limit of the large number of colors $N_c$. In the the  multicolor limit only nearest neighbor interactions are not suppressed and  the BFKL dynamics is similar to that of the Ising spin chain model.

The transformation Eq.~(\ref{change}) of the holomorphic BFKL Hamiltonian is an unitary transformation only for a vanishing total momentum
\begin{eqnarray}
 \vec{p}=\sum^n_{r=1}\vec{p}_r
\end{eqnarray}
which guarantees the cyclicity of the momenta $p_r$ important for their representation by the difference of coordinates $\rho_{r,r+1}$. For the compound state of two reggeized gluons~(usual BFKL case) for $n=2$, this can be achieved only for the zero transferred momentum $\vec{q}=0$. Only in this case one can really identify the dual coordinates  $\vec{\rho}_{r,r+1}$ of the momenta $\vec{p}_r$ with their conjugate coordinates. In a more general case these two are not the same object. However, the integrability of the non-forward BFKL suggests that the duality symmetry should be present also in the case of $\vec{q}\neq 0$, but in an \emph{implicit} way. The main objective of the present study is to show that the dipole formulation of the BFKL evolution can provide a suitable framework for studying the duality symmetry of the non-forward BFKL. We show that the evolution equation for the scattering of a non-diagonal dipole coincides with the non-forward BFKL equation in the dual space provided we impose on the dipole scattering amplitude some condition that is dual to the so-called \emph{BFKL condition}. The  BFKL condition is a result of the unitarity and the multiregge kinematics used for deriving the leading order BFKL as discussed below.  In this formalism the duality symmetry of the non-forward BFKL equation appears in an implicit way due to the fact that we can identify the set of coordinates of the scatterred dipole with a set of dual coordinates of the reggeized gluons momenta. However it looks like that there is no  obvious  choice of the  Fourier transform that can relate the dipole coordinates to the reggeized gluon momenta individually, that is the reason why the duality symmetry is established implicitly. The duality symmetry holds also in the non-forward case because of the special structure of the non-forward BFKL Kernel, which can be viewed as sum of the three forward Kernels. As it was already mentioned the duality symmetry of the forward BFKL can be shown explicitly, which suggests that the sum of the three forward Kernels should also possess this property.  

One remark is in order. A system may possess another symmetry with a similar name, called the \emph{dual conformal} symmetry. 
The dual conformal symmetry is an usual conformal symmetry in dual coordinates~($k_i=x_i-x_{i+1}$) and, generally, is not related to the duality symmetry. 
The dual conformal symmetry is now successfully implemented  in calculating multileg planar amplitudes in SYM $\mathcal{N}=4$~(for an up-to-date discussion see Ref.~\cite{Korchemsky:2009hm} and references wherein), and it was also recently considered in the connection with the BFKL equation~\cite{Gomez:2009bx}. 
This symmetry is beyond the scope of the present study.

In the next section we write the BFKL equation in the dual coordinates and analyze its structure. We argue that the non-forward BFKL equation can be represented as a three point amplitude, due to the \emph{BFKL condition} associated with the lack of the crossing symmetry in the BFKL approach.

\section{BFKL equation in dual coordinates}
\label{sec:BFKLdualcoordinates}
\hspace{0.4cm} In this section we discuss the structure of the BFKL equation and write it in the \emph{dual} coordinates. We show that the \emph{non-forward} BFKL Kernel $q\neq 0$ can written as a sum of three \emph{forward} Kernels, which can be interpreted as two uncut and one cut Kernel~(UCU structure). The UCU structure of the BFKL equation is crucial for establishing the duality symmetry also in the non-forward case. 

We start with recasting the BFKL equation into a form useful for our discussion.\footnote{From now on we deal only with two dimensional transverse momenta and omit the vector sign to make the presentation clear.} The direct substitution of Eq.~(\ref{regge}) and Eq.~(\ref{kernel}) in 
Eq.~(\ref{bfkl}) gives 
\begin{eqnarray}\label{bfkl3linesOLD}
  \frac{\partial \mathcal{F}(k,k-q) }{\partial y}&=&+ 
\frac{\alpha_s N_c}{2 \pi^2} \int  \frac{d^2 \chi \;k^2}{ \chi^2(\chi-k)^2}\mathcal{F}(\chi,\chi-q)-\frac{\alpha_s N_c}{4 \pi^2} \int   \frac{d^2 \chi \;k^2}{ \chi^2(\chi-k)^2}\mathcal{F}(k,k-q) \nonumber \\
&&
+\frac{\alpha_s N_c}{2 \pi^2} \int   \frac{d^2 \chi \; (k-q)^2}{ (\chi-q)^2(\chi-k)^2}\mathcal{F}(\chi,\chi-q)-\frac{\alpha_s N_c}{4 \pi^2} \int   \frac{d^2 \chi \;(k-q)^2}{ (\chi-q)^2(\chi-k)^2}\mathcal{F}(k,k-q)   \nonumber \\
&&
-\frac{\alpha_s N_c}{2 \pi^2} \int   \frac{d^2 \chi \; q^2}{ \chi^2(\chi-q)^2}\mathcal{F}(\chi,\chi-q)
\end{eqnarray}

For our purpose it is convenient to write the second line of  Eq.~(\ref{bfkl3linesOLD}) in a slightly different form changing the integration variable $\chi \to \chi-q$ 
\begin{eqnarray}\label{bfkl3lines}
  \frac{\partial \mathcal{F}(k,k-q) }{\partial y}&=&+ 
\frac{\alpha_s N_c}{2 \pi^2} \int  \frac{d^2 \chi \;k^2}{ \chi^2(\chi-k)^2}\mathcal{F}(\chi,\chi-q)-\frac{\alpha_s N_c}{4 \pi^2} \int   \frac{d^2 \chi \;k^2}{ \chi^2(\chi-k)^2}\mathcal{F}(k,k-q) \nonumber \\
&&
+\frac{\alpha_s N_c}{2 \pi^2} \int   \frac{d^2 \chi \; (k-q)^2}{ \chi^2(\chi-k+q)^2}\mathcal{F}(\chi+q,\chi)-\frac{\alpha_s N_c}{4 \pi^2} \int   \frac{d^2 \chi \;(k-q)^2}{ \chi^2(\chi-k+q)^2}\mathcal{F}(k,k-q)   \nonumber \\
&&
-\frac{\alpha_s N_c}{2 \pi^2} \int   \frac{d^2 \chi \; q^2}{ \chi^2(\chi-q)^2}\mathcal{F}(\chi,\chi-q)
\end{eqnarray}

\begin{figure}[h]
\begin{center}
\includegraphics[width=4.5in]{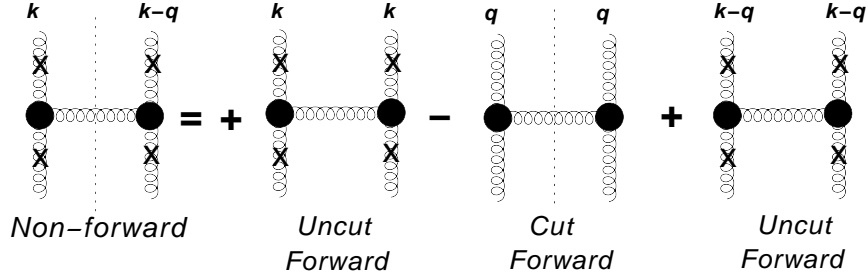}  
\end{center}
\caption{ The uncut-cut-uncut~(UCU) structure of the non-forward BFKL. The non-forward BFKL Kernel can be written as a linear combination of three forward Kernels, two uncut~(for two gluon pairs $k$ and $k-q$) and one cut for the gluon pair $q$. The cut Kernel does not possess virtual contributions, which is reflected by the absence of crosses on gluons $q$.} \label{fig:UCUnew}
\end{figure}

One can see that the Kernel of the non-forward BFKL equation can be written as a sum of the three forward Kernels, where two of them, are the usual forward BFKL Kernels given by the first and the second lines of Eq.~(\ref{bfkl3lines}), while the third line has only real emission part. This interpretation is better understood from Fig.~\ref{fig:UCUnew}, where the unitarity cut is denoted by the vertical dashed line. The BFKL Kernel that describes the bound states of two reggeized gluons $k$ and $k-q$ can be viewed as sum of the  \emph{uncut} forward Kernels for the gluon pair $k$ and $k$ and the gluon pair $k-q$ and $k-q$  ( the first term and the third term on the r.h.s in Fig.~\ref{fig:UCUnew} ),  and the \emph{cut} forward Kernel for the scattering of the pair of fictitious gluons $q$ and $q$ (the second term on the r.h.s in  Fig.~\ref{fig:UCUnew}  ). The last contribution seats exactly on the unitarity cut and thus does not possess any virtual contribution. This UCU structure of the non-forward BFKL Kernel plays an important role in showing the duality  symmetry of the BFKL equation and in finding its dual in the dipole picture as we show below. To see this we first write Eq.~(\ref{bfkl3lines}) in the  \emph{dual} coordinates properly chosen by making the following observations.

The duality symmetry holds also for a case of the multireggeon exchange, and is not limited to the system of two reggeized gluons as in the BFKL equation Eq.~(\ref{bfkl3lines}). Another important point is that only the upper momenta of the reggeized gluons are to be taken into account. In particular, this means that in the case of the BFKL Pomeron we have only three momenta for the duality symmetry, because the Regge kinematics selects $t$-channel for a propagation of the BFKL state breaking the crossing symmetry. Together with the unitarity condition and the strong ordering of the produced particles~(multiregge kinematics) this results into some constraint on the form of the leading order BFKL amplitude, which we call the $BFKL \hspace{0.1cm} condition$. This condition is implicitly written in the LO BFKL as we explain below.

By inspecting the arguments of the BFKL amplitude in Eq.~(\ref{bfkl3lines}) of both the real and the virtual parts, one can deduce that their difference is always equal to the transferred momentum, namely for $\mathcal{F}(k_i,k_j)$ we have $k_i-k_j=q$. This is a consequence of the use of the $t$-channel unitarity together with a special kinematics in the BFKL approach. We call this condition the $BFKL \hspace{0.1cm} condition$. It suggests to treat the BFKL amplitude as a three point function with external momenta

\begin{eqnarray}\label{k1k2k3}
k_1=k ;\;\; k_2=q-k; \;\; k_3=-q
\end{eqnarray}

At first sight, the BFKL amplitude is four point scattering amplitude with four external (transverse)~momenta $k$, $k-q$, $k'$ and $k'-q$, but the BFKL condition removes the necessity in the fourth external momentum leaving three momenta which obey the conservation law.
 This means that the BFKL amplitude is in fact a function of only two external transverse momenta, i.e. $k$ and $q$ or $k$ and $k-q$. In other words, the duality symmetry deals  with only upper gluon momenta or only lower gluon momenta, but never with mixes them.  This observation suggests to pick up only three dual coordinates.

\begin{figure}[h]
\begin{center}
\hspace{-1cm}\includegraphics[width=6in]{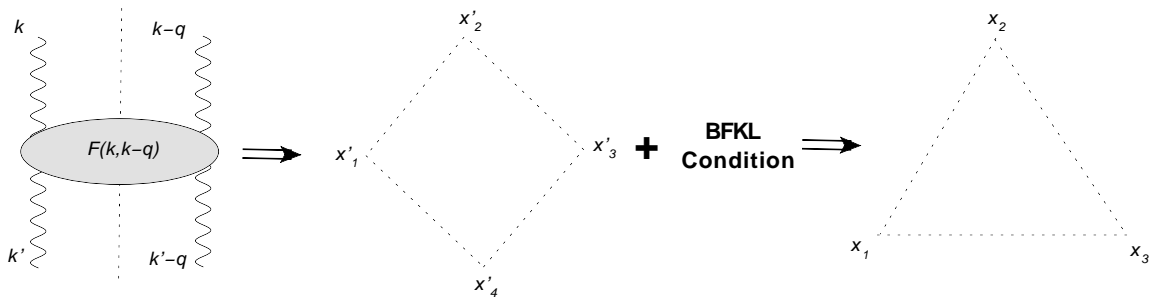}  
\end{center}
\caption{The BFKL condition constraints the representation of the BFKL scattering amplitude as a function of only three dual coordinates instead of four external points.  } \label{fig:4point3point}
\end{figure}

For our purposes we define the dual coordinates 

\begin{eqnarray}\label{duals}
 k=k_1=x_1-x_2=x_{12}; \;\;  q-k=k_2=x_2-x_3=x_{23}; \;\; -q=k_3=x_3-x_1=x_{31}
\end{eqnarray}
so that the overall momenta conservation $k_1+k_2+k_3=0$ is automatically satisfied and the BFKL amplitude can be represented as a three point function in the dual space as illustrated in Fig.~\ref{fig:4point3point}. Using this definition we can write the BFKL equation Eq.~(\ref{bfkl3lines}) as follows

 \begin{eqnarray}\label{bfkldual}
  \frac{\partial \mathcal{F}(x_{12},x_{23}) }{\partial y}&=&+ 
\frac{\alpha_s N_c}{2 \pi^2} \int  \frac{d^2 z \; x_{12}^2}{ z^2(z-x_{12})^2}\left\{\mathcal{F}(z,z+x_{31})-\frac{1}{2}\mathcal{F}(x_{12},-x_{23}) \right\} \nonumber \\
&&
+\frac{\alpha_s N_c}{2 \pi^2} \int   \frac{d^2 z \; x_{23}^2}{ z^2(z+x_{23})^2}\left\{\mathcal{F}(z-x_{31},z)-\frac{1}{2}\mathcal{F}(x_{12},-x_{23}) \right\}  \nonumber \\
&&
-\frac{\alpha_s N_c}{2 \pi^2} \int   \frac{d^2 z \; x_{31}^2}{ z^2(z+x_{31})^2}\mathcal{F}(z,z+x_{31})
\end{eqnarray}

In the next section we discuss the evolution equation for the non-diagonal dipole scattering and show that it can be written in the form of the BFKL in the dual coordinates  Eq.~(\ref{bfkldual}) by imposing on it the condition dual to the BFKL condition.

\section{Scattering of non-diagonal dipole}
\label{sec:dipole}

\begin{figure}[h]
\begin{center}
\hspace{-1cm}\includegraphics[width=2in]{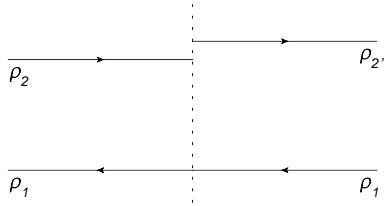}  
\end{center}
\caption{ The schematic representation of a color dipole, which has different transverse sizes to the left and to the right of the unitarity cut denoted by the vertical dashed line. For our purposes it is enough to consider only difference in the coordinates of the upper~(antiquark) line, keeping the the coordinates of the lower~(quark) the same. The broken antiquark line  illustrates  only the fact that the sizes are different. There is no discontinuity in the charge flow etc. } \label{fig:dipole12psik}
\end{figure}

\hspace{0.4cm} In this section we show that the evolution equation for the scattering of the non-diagonal dipole depicted in Fig.~\ref{fig:dipole12psik} can be brought to the form of the BFKL equation in the dual space Eq.~(\ref{bfkldual}). The scattering of the non-diagonal dipole with different coordinates in the amplitude and the conjugate amplitude (to the left and to the right of the unitarity cut)  was considered by Levin and the author \cite{Levin:1900tt} as an auxiliary problem in proving the single inclusive production formula in the dipole formulation.
Such a dipole can be constructed if one fixes momentum of the antiquark line and thus keeping it coordinates different to the left and to the right of the unitarity cut, whereas the lower quark line momentum is integrated over resulting in $\delta^{(2)}(\rho_1-\rho_{1'})$.
 The non-linear evolution equation was derived and solved using the notion of the "generalized optical theorem''. The function $M(12|12')$ for which the equation was derived is an auxiliary function for proving the single inclusive production formula for the dipole model. It has a meaning of the non-diagonal dipole total cross section since for $\rho_2=\rho_{2'}$ it reduces to $M(12|12)=2N(12)$ (the optical theorem in the coordinate space), where $N(12)$ is the BFKL amplitude in the M\"obius representation. This property was imposed by the definition of $M(12|12')$, since for $\rho_2=\rho_{2'}$ the non-diagonal dipole takes a form of a usual dipole, which is described by the BFKL equation.

The evolution equation for the non-diagonal dipole is derived using \emph{real}-\emph{virtual} \emph{non}-\emph{cancellations}, i.e. including the interactions in the final state.  The final state interactions fully cancel in the inclusive case, but as far as gluon production is concerned such cancellation does not happen and this fact is crucial for obtaining the closed form of the single gluon production cross section with evolution effects included. For more details about the way it was derived one is referred to Levin and the author~\cite{Levin:1900tt}. Here we only want  to discuss the linear version of this evolution equation, its properties and to show that it can be written as a non-forward BFKL in the dual space. This result would mean that  there exist  a \emph{hidden} \emph{duality symmetry}  of the non-forward BFKL, that appears implicitly from our analysis due to the fact that the set of  dual coordinates~(with dimensions of mass) can be associated with set of the transverse coordinates of the dipoles. This extends  the duality symmetry  shown by Lipatov for the forward case, to a non-zero momentum transfer, which can potentially explain the integrability of the BFKL equation. 

 For our purposes we retain only the linear part of the resulting non-linear evolution equation for a non-diagonal dipole scattering~\cite{Levin:1900tt}. It reads

\begin{eqnarray}\label{evolMA1simple}
\frac{\partial M(12|12')}{\partial y}=\frac{\bar{\alpha_s}}{2\pi}\int d^2\rho_3 \left\{
-\frac{1}{2}\left(\frac{\rho_{13}}{\rho^2_{13}}-\frac{\rho_{23}}{\rho^2_{23}}\right)^2M(12|12')-\frac{1}{2}\left(\frac{\rho_{13}}{\rho^2_{13}}-\frac{\rho_{2'3}}{\rho^2_{2'3}}\right)^2M(12|12') 
\right.
\end{eqnarray}
\begin{eqnarray} 
\left.
+\left(\frac{\rho_{13}}{\rho^2_{13}}-\frac{\rho_{23}}{\rho^2_{23}}\right)\left(\frac{\rho_{13}}{\rho^2_{13}}-\frac{\rho_{2'3}}{\rho^2_{2'3}}\right) \left\{2N(13)+M(32|32')\right\}
-\left(\frac{\rho_{13}}{\rho^2_{13}}-\frac{\rho_{23}}{\rho^2_{23}}\right)\left(\frac{\rho_{23}}{\rho^2_{23}}-\frac{\rho_{2'3}}{\rho^2_{2'3}}\right)
M(13|12')\right. \nonumber
\\
\left.
-\left(\frac{\rho_{2'3}}{\rho^2_{2'3}}-\frac{\rho_{23}}{\rho^2_{23}}\right)\left(\frac{\rho_{13}}{\rho^2_{13}}-\frac{\rho_{2'3}}{\rho^2_{2'3}}\right)
M(12|13)
-\frac{1}{2}\left(\frac{\rho_{23}}{\rho^2_{23}}-\frac{\rho_{2'3}}{\rho^2_{2'3}}\right)^2
M(12|12') 
\right\}  \nonumber
\end{eqnarray}

As it was already mentioned the function $M(12|12')$ is defined such that  $M(12|12)=2N(12)$.\footnote{Here indices of the argument stand for the transverse coordinates of the quark $\rho_1$ and the antiquark $\rho_{2}$~($\rho_{2'}$) lines and not for only  the dipole size $\rho_{12}=\rho_1-\rho_2$ in contrast to the common notation.} This definition follows from the fact that in the simple case of equal dipole sizes $\rho_2=\rho_{2'}$ all necessary \emph{real-virtual cancellations} take place removing all final state interactions, and one deals with the  scattering of an usual color dipole described by the BFKL equation in the coordinate space. 
Indeed, as it easy to see that Eq.~(\ref{evolMA1simple})  reduces to the BFKL equation for $\rho_2=\rho_{2'}$~(see Ref.~\cite{Levin:1900tt}).
This definition and the properties of the initial condition suggested the possible form of the solution to the non-diagonal dipole evolution equation
\begin{eqnarray}\label{optics}
 M(12|12')=N(12)+N(12')-N(22')
\end{eqnarray}
It was checked by the explicit substitution that this form of the solution keeps also in the non-linear case of the generalized Balitsky-Kovchegov~(BK) \cite{B,K} equation  considered in Ref.~\cite{Levin:1900tt}. The non-linear equation is a generalization of the Balitsky-Kovchegov equation and coincides with it for $\rho_2=\rho_{2'}$ similar to the linear case.
Using this form of  solution in  Eq.~(\ref{evolMA1simple}) we obtain

\begin{eqnarray}\label{evolMA1comb}
\frac{\partial (N(12)+N(12')-N(22'))}{\partial y}=\frac{\bar{\alpha_s}}{2\pi}\int 
 \frac{\rho^2_{12}d^2\rho_3 }{\rho^2_{13}\rho^2_{23}}  \left\{N(13)+N(32)-N(12)\right\}
 \end{eqnarray}
\begin{eqnarray} 
+\frac{\bar{\alpha_s}}{2\pi}\int 
 \frac{\rho^2_{12'}d^2\rho_3 }{\rho^2_{13}\rho^2_{2'3}} \left\{N(13)+N(32')-N(12')\right\} \nonumber
\end{eqnarray}
\begin{eqnarray} 
-\frac{\bar{\alpha_s}}{2\pi}\int 
 \frac{\rho^2_{22'}d^2\rho_3 }{\rho^2_{23}\rho^2_{2'3}}  \left\{N(32)+N(32')-N(22')\right\} \nonumber
\end{eqnarray}

which is just a linear combination of three BFKL equations for initial dipoles with coordinates $12$, $12'$ and $22'$.  This reminds the \emph{uncut-cut-uncut} ~(UCU) structure of the BFKL equation in the momentum space mentioned in the previous section.
It is worthwhile mentioning that generalized BK equation also has the UCU structure, which fully corresponds to the picture drawn by Ciafaloni, Marchesini and Veneziano deriving the Cut Reggeon Calculus \cite{Ciafaloni:1975jy,Ciafaloni:1975fh}. They found that the Pomeron can be described as a linear combination of three propagating states, which correspond to one cut and two uncut Pomerons $\phi^++\phi^--\phi^c$. The reggeon field $\phi^+$ stands for the Pomeron to the left of the unitarity cut, $\phi^-$ for the Pomeron to the right of the unitarity cut and $\phi^c$ represents the Pomeron living on the cut.   The introduction of the triple Pomeron splitting vertex~(``fan`` diagrams) preserves this structure, while the Pomeron loops break it explicitly. The same result was also obtained by Levin and the author \cite{Levin:2007yv} using generating functional approach to the analysis of the multiparticle states in the dipole model based on Abramovski-Gribov-Kancheli cutting rules~\cite{Abramovsky:1973fm}.

Our goal is to show that the evolution equation for the non-diagonal dipole reproduces the non-forward BFKL equation in the dual coordinates. It is not difficult to see that with the help of the solution  Eq.~(\ref{optics}) we can recast Eq.~(\ref{evolMA1comb}) into form of

\begin{eqnarray}\label{evolMnew1}
\frac{\partial M(12|12')}{\partial y}&=&\frac{\bar{\alpha_s}}{2\pi}\int 
 \frac{\rho^2_{12}d^2\rho_3 }{\rho^2_{13}\rho^2_{23}}  \left\{M(32|32')-\frac{1}{2}M(12|12')+M(32|22')-\frac{1}{2}M(12|22')\right\} 
\\&+&
\frac{\bar{\alpha_s}}{2\pi}\int 
 \frac{\rho^2_{12'}d^2\rho_3 }{\rho^2_{13}\rho^2_{2'3}} \left\{M(32|32')-\frac{1}{2}M(12|12')+M(32'|22')-\frac{1}{2}M(12'|22')\right\}\nonumber
\\&-&
\frac{\bar{\alpha_s}}{2\pi}\int 
 \frac{\rho^2_{22'}d^2\rho_3 }{\rho^2_{23}\rho^2_{2'3}}  M(32|32') \nonumber
\end{eqnarray}

We immediately notice that Eq.~(\ref{evolMnew1}) is very similar to Eq.~(\ref{bfkldual}) except the last two terms in brackets of the first two lines. This is despite the fact that the functions $M(12|12')$ is defined in a very much different way than the BFKL amplitude. The main difference is that the BFKL amplitude $\mathcal{F}(k,k-q)$ accounts for the requirement of the BFKL condition. In particular, this means that for $\mathcal{F}(k,k-q)=\mathcal{F}(k_1,k_2)$ the arguments should satisfy 
\begin{eqnarray}\label{unit}
k_1-k_2=q
\end{eqnarray}

which is translated in terms of the  dual coordinates Eq.~(\ref{duals}) as the function $\mathcal{F}(z_1,z_2)$ should satisfy $z_1-z_2=-x_{31}$. It  is now  possible to identify the dual coordinates of Eq.~(\ref{duals}) with the transverse dipole coordinates as follows

\begin{eqnarray}\label{associate}
 \rho_{12}=x_{12}; \;\; \rho_{12'}=-x_{23}; \;\; \rho_{22'}=x_{31}
\end{eqnarray}

The dual of the BFKL condition in Eq.~(\ref{unit}) for $M(ij|ik)$ reads 
\begin{eqnarray}\label{unitdual}
\rho_{ij}-\rho_{ik}=-\rho_{22'}
\end{eqnarray}
Imposing the dual of the BFKL condition Eq.~(\ref{unitdual}) on the evolution equation for the non-diagonal dipole removes undesired terms in Eq.~(\ref{evolMnew1}) and we are left with 

\begin{eqnarray}\label{dipoledualOLD}
\frac{\partial \tilde{M}(\rho_{12}|\rho_{12'})}{\partial y}&=&\frac{\bar{\alpha_s}}{2\pi}\int 
 \frac{\rho^2_{12}d^2\rho_3 }{\rho^2_{13}\rho^2_{23}}  \left\{\tilde{M}(\rho_{32}|\rho_{32'})-\frac{1}{2}\tilde{M}(\rho_{12}|\rho_{12'})\right\}
\\&+&
\frac{\bar{\alpha_s}}{2\pi}\int 
 \frac{\rho^2_{12'}d^2\rho_3 }{\rho^2_{13}\rho^2_{2'3}}\left\{\tilde{M}(\rho_{32}|\rho_{32'})-\frac{1}{2}\tilde{M}(\rho_{12}|\rho_{12'})\right\} \nonumber
\\&-&
\frac{\bar{\alpha_s}}{2\pi}\int 
 \frac{\rho^2_{22'}d^2\rho_3 }{\rho^2_{23}\rho^2_{2'3}} \tilde{M}(\rho_{32}|\rho_{32'}) \nonumber
\end{eqnarray}

Recasting Eq.~(\ref{dipoledualOLD}) in a more transparent form we get

\begin{eqnarray}\label{dipoledualOLD}
\frac{\partial \tilde{M}(\rho_{12}|\rho_{12'})}{\partial y}&=&\frac{\bar{\alpha_s}}{2\pi}\int 
 \frac{\rho^2_{12}d^2\rho_{32} }{\rho^2_{32}(\rho_{32}-\rho_{12})^2}  \left\{\tilde{M}(\rho_{32}|\rho_{32}+\rho_{22'})-\frac{1}{2}\tilde{M}(\rho_{12}|\rho_{12'})\right\}
\\&+&
\frac{\bar{\alpha_s}}{2\pi}\int 
 \frac{\rho^2_{12'}d^2\rho_{32'} }{\rho^2_{32'}(\rho_{32'}-\rho_{12'})^2}\left\{\tilde{M}(\rho_{32'}-\rho_{22'}|\rho_{32'})-\frac{1}{2}\tilde{M}(\rho_{12}|\rho_{12'})\right\} \nonumber
\\&-&
\frac{\bar{\alpha_s}}{2\pi}\int 
 \frac{\rho^2_{22'}d^2\rho_{32} }{\rho^2_{32}(\rho_{32}+\rho_{22'})^2} \tilde{M}(\rho_{32}|\rho_{32}+\rho_{22'}) \nonumber
\end{eqnarray}

which is identical to the non-forward BFKL equation in the dual space Eq.~(\ref{bfkldual}) provided we identify the dipole coordinates and the dual coordinates as in Eq.~(\ref{associate}). 
 It is not surprising that the equation for $M(12|12')$ includes more terms than the BFKL for $\mathcal{F}(k,k-q)$, since $M(12|12')$ was defined without any additional constraint except to reproduce dipole BFKL for $\rho_2=\rho_{2'}$, in contrast to the BFKL amplitude.

By construction of the dipole model the  coordinates $\rho_{ij}$ are conjugate to the momenta $k_i$ of the reggeized gluons. The fact that we can identify the dual momenta coordinates of Eq.~(\ref{duals}) with the dipole coordinates indicates that the duality symmetry is preserved also in the non-forward case. However, there seems to be no obvious way to introduce the Fourier transform that connects them and thus  the duality symmetry is \emph{hidden}.

In our discussion we ignored the issue of the initial condition and the impact parameter $b_{12}=(\rho_1+\rho_2)/2$ dependence.
These two are related to each other since the impact parameter defines a reference point that connects the evolution to the target. Any fixed reference point breaks explicitly the translational symmetry and thus the impact parameter cannot be related to the set of the dual coordinates Eq.~(\ref{duals}). For a similar  reason we do not consider the lower momenta $k'$ and $k'-q$ of the amputated BFKL amplitude shown in Fig.~\ref{fig:4point3point}, more accurately, we do not consider simultaneously the upper and the lower momenta. We assign the evolution to the upper momenta, while the lower momenta enter through the initial condition~(we could do vice versa). Any attempt to include the initial condition to duality picture would contradict the lack of the impact parameter dependence in the dipole picture, but as have already pointed out the $b$-dependence is incompatible with the requirement of the translational invariance.

\begin{figure}[h]
\begin{center}
\hspace{-1cm}\includegraphics[width=4in]{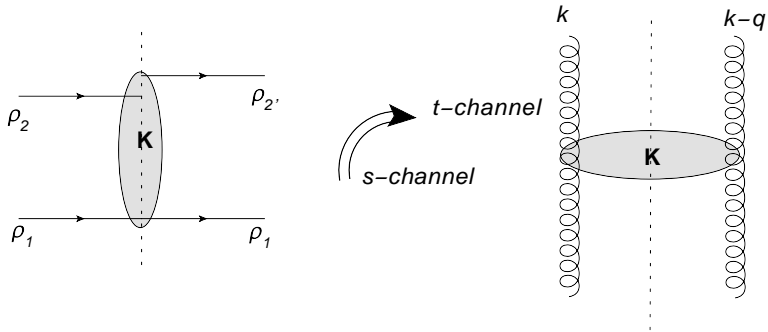}  
\end{center}
\caption{The duality symmetry can be interpreted as a symmetry under rotation of the BFKL Kernel in the transverse space from $s$-channel~(color dipoles) to $t$-channel~(reggeized gluons). The unitarity cut is denoted by a dashed vertical line.} \label{fig:rot}
\end{figure}

The hidden duality symmetry is related only to the pure evolution, without any reference to the initial condition. As it was anticipated in Ref.~\cite{Prygarin:2009tn}, the duality symmetry is the symmetry under rotation of the BFKL Kernel in the transverse space from  $s$-channel to $t$-channel and back as illustrated in Fig.~\ref{fig:rot}. This rotation is, in fact, a rotation between the reggeized gluon formulation of the BFKL evolution and  the dipole picture. The connection between the two pictures is certainly not complete without matching the initial condition. The proper matching is formulated as follows. At the first stage, one makes a suitable choice of the dual coordinates, then the physical picture is changed by rotating the Kernel of the evolution equation in the \emph{transverse} space and the function is given the proper interpretation~(either reggeized gluon or dipole scattering amplitude). Finally, at the second stage, the initial conditions are chosen in accordance to the physical picture. The second stage is obviously has nothing to do with duality symmetry property of the BFKL evolution. This point seems to be not so much important in the case of the linear evolution considered here, but it becomes crucial for clearifying the meaning of the duality symmetry of the Balitsky-Kovchegov equation.

\section{Conclusion}
\label{sec:concl}

\hspace{0.4cm} We discussed the duality symmetry of the LO BFKL equation. The duality symmetry of the BFKL equation was formulated  by Lipatov~\cite{Lipatov:1998as} for a system of $n$ reggeized gluons, and in the case of the color singlet BFKL equation~($n=2$) the duality symmetry  was shown to hold only in the forward~($q=0$) case. In the present study we argue that the duality symmetry is valid also in the non-forward case, though in an \emph{implicit} way. The \emph{hidden} duality symmetry is established by identifying the dual coordinates~(with dimension of mass) of the BFKL in the momentum space with the transverse sizes of a non-diagonal dipole scattered off the target.  The evolution equation for the non-diagonal dipole having different sizes to the right and to the left of the unitarity cut was derived by Levin and the author~\cite{Levin:1900tt}. Its analytical solution was also found, and it is a linear combination of three amplitudes of usual dipoles. This structure is similar to the structure of the non-forward BFKL, which  can  be also decomposed in three pieces each corresponding to forward BFKL. Two of the pieces can be viewed as uncut BFKL, while one piece does not have virtual contribution and is interpreted as a cut BFKL. The uncut-cut-uncut~(UCU) structure of the BFKL Kernel uncovered in the present study is consistent with the picture drawn by Ciafaloni, Marchesini and Veneziano~\cite{Ciafaloni:1975fh,Ciafaloni:1975jy} in Cut Reggeon Calculus, where the Pomeron is represented by three fields, which denote two uncut and one cut Pomerons.

We argue that  the duality symmetry can be viewed as a symmetry under rotation of the BFKL Kernel in the transverse space from $s$-channel~(color dipoles) to $t$-channel~(reggeized gluons) and back as illustrated in Fig.~\ref{fig:rot}.

\subsection*{Acknowledgments}
\hspace{0.4cm}
We are deeply indebted to J.~Bartels, V.~Fadin, G.~Korchemsky, J.~Kotanski, L.~Levin, L.~Lipatov and L.~Motyka for very helpful discussions.  This study was supported by the Minerva Postdoctoral Fellowship of the Max Planck Society.

\newpage
 \hspace{0.5cm}

\end{document}